\documentclass[epj]{webofc}
\usepackage[varg]{txfonts}   
\woctitle{Fifth International Workshop on Nuclear fission and Fission-Product Spectroscopy }
\begin{document}
\title{Nuclear fission in covariant density functional theory}
%
%

\author{A.\ V.\ Afanasjev\inst{1}\fnsep\thanks{\email{afansjev@erc.msstate.edu}},
        H.\ Abusara\inst{2}, 
        P.\ Ring\inst{3}
}

\institute{Department of Physics and Astronomy, Mississippi State University, MS 39762, USA
\and
   Department of Physics, Faculty of Science, An-Najah National University, Nablus, Palestine
\and
           Fakult\"at f\"ur Physik, Technische Universit\"at M\"unchen,
 D-85748 Garching, Germany}

\abstract{%
The current status of the application of covariant density functional
theory to microscopic description of nuclear fission with main emphasis
on superheavy nuclei (SHN) is reviewed. The softness of SHN in the 
triaxial plane leads to an emergence of several competing fission pathes 
in the region of the inner fission barrier in some of these nuclei. The 
outer fission barriers of SHN are considerably affected both by 
triaxiality and octupole deformation.}
\maketitle
\section{Introduction}
\label{intro}

 A study of the fission barriers of nuclei is motivated by their
importance for several physical phenomena. For example, the 
$r-$process of stellar nucleosynthesis
depends (among other quantities such as masses and $\beta$-decay
rates) on the fission barriers of very neutron-rich nuclei
\cite{MPR.01}. The population and survival
of hyperdeformed states at high spin also depends on the fission
barriers \cite{AA.08}. In addition, the physics of fission 
barriers is intimately connected with on-going search for new 
superheavy nuclei (SHN). 
The probability for the formation of a SHN in a heavy-ion-fusion reaction
is directly connected to the height of its fission barrier \cite{IOZ.02};
the large sensitivity of the cross section $\sigma$ for the synthesis of
the fissioning nuclei on the barrier height $B_{f}$ also stresses
a need for accurate calculations of this value. The survival of the
actinides and SHN against spontaneous fission depends on the fission
barrier which is a measure of the stability of a nucleus reflected in
the spontaneous fission lifetimes of these nuclei \cite{SP.07}. The 
recent progress in the microscopic description of fission barriers within 
covariant density functional theory (CDFT) \cite{VALR.05} is briefly 
reviewed in the current manuscript with the main emphasis on SHN.

\section{Comparison with other models}
\label{sec}

\begin{figure*}[ht]
\begin{center}
\includegraphics[width=11.0cm,angle=0]{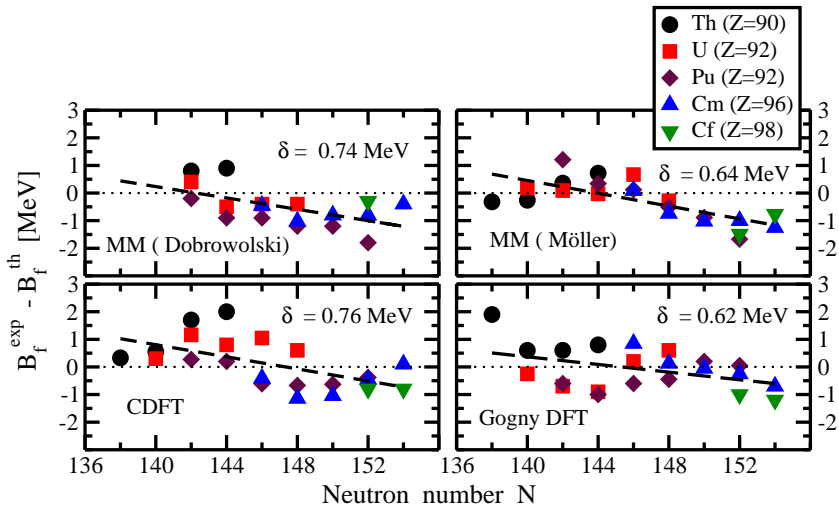}
\end{center}
\vspace{-0.6cm}
\caption{The difference between experimental and calculated heights
of inner fission barriers as a function of neutron number $N$. The 
results of the calculations are compared to estimated fission barrier heights 
given in the RIPL-3 database \cite{RIPL-3}.
The results of the calculations within the microscopic+macroscopic
method ('MM(Dobrowolski)' \cite{DPB.07} and 'MM(M{\"o}ller)' \cite{MSI.09}),
covariant density functional theory ('CDFT' \cite{AAR.10}) and density
functional theory based on the finite range Gogny D1S force ('Gogny DFT'
\cite{DGGL.06}) are shown. 
The average deviation per barrier $\delta$ [in MeV] is defined as
$\delta = \sum_{i=1}^N |B_f^i(th)-B_f^i(exp)|/N$, where $N$ is the number
of the barriers with known experimental heights, and $B_f^i(th)$
($B_f^i(exp)$) are calculated (experimental) heights of the barriers.
Long-dashed lines represent the trend of the deviations between theory and
experiment as a function of neutron number. 
From Ref.\ \cite{AAR.12}.}
\label{N-dep}
\end{figure*}
  The progress in the development of computer codes and the availability of 
powerful computers has allowed to study in a systematic way the effects of 
triaxiality on the fission barriers leading to their realistic description. 
Within the CDFT framework, the inner fission barriers with triaxiality 
included have been studied for the first time in Ref.\ \cite{AAR.10} using the 
triaxial RMF+BCS approach and the NL3* parametrization \cite{NL3*}. Two years 
later, similar studies of fission barriers in actinides have been performed
in Refs.\ \cite{LZZ.12,PNLV.12} using the RMF+BCS framework with the PC-PK1 
\cite{PC-PK1} and DD-PC1 \cite{DD-PC1} CDFT parametrizations. 
The accuracy of the description of the heights of inner fission barriers in 
these calculations is comparable with the one obtained in Ref.\ \cite{AAR.10}. 
The calculations of Refs.\ \cite{LZZ.12,PNLV.12} also include the results for 
outer fission barriers where the effects of octupole deformation (and 
triaxiality [Ref.\ \cite{LZZ.12}]) are taken into account. They agree well 
with experimental data.

  The inclusion of triaxiality has drastically improved the accuracy 
of the description of inner fission barriers in all model calculations 
(Ref.\ \cite{AAR.12}). The common 
consensus is that reflection asymmetric (octupole) deformations do
not affect inner fission barriers, but have considerable impact on the
outer fission barriers in actinides and SHN. Fig.\
\ref{N-dep}\footnote{Similar figure but as a function of proton number 
$Z$ is shown in Ref.\ \cite{AAR.12}.} shows that the state-of-the-art 
calculations within different theoretical frameworks (including CDFT) 
are characterized by a comparable accuracy (the $\delta$-values) of 
the description of inner fission barriers. Good description of 
inner fission barriers has also been obtained in the microscopic+macroscopic
(MM) calculations of Ref.\ \cite{KJS.10}; however, these calculations 
do not include Th isotopes. Recent Skyrme DFT calculations are also 
characterized by similar accuracy of the description of inner fission 
barriers \cite{SDFT.12}. The energies of fission isomers and outer 
fission barriers are also described with the accuracy similar to the one 
of inner fission barriers in the above mentioned model calculations 
\cite{SDFT}. Minor differences between the approaches in the 
obtained average deviations per barrier (Fig.\ \ref{N-dep}) are not 
important considering appreciable uncertainties in the extraction 
of fission barrier heights from experimental data (see Ref.\ 
\cite{AAR.12-int}).

\begin{figure*}[ht]
\begin{center}
\includegraphics[width=7.0cm,angle=0]{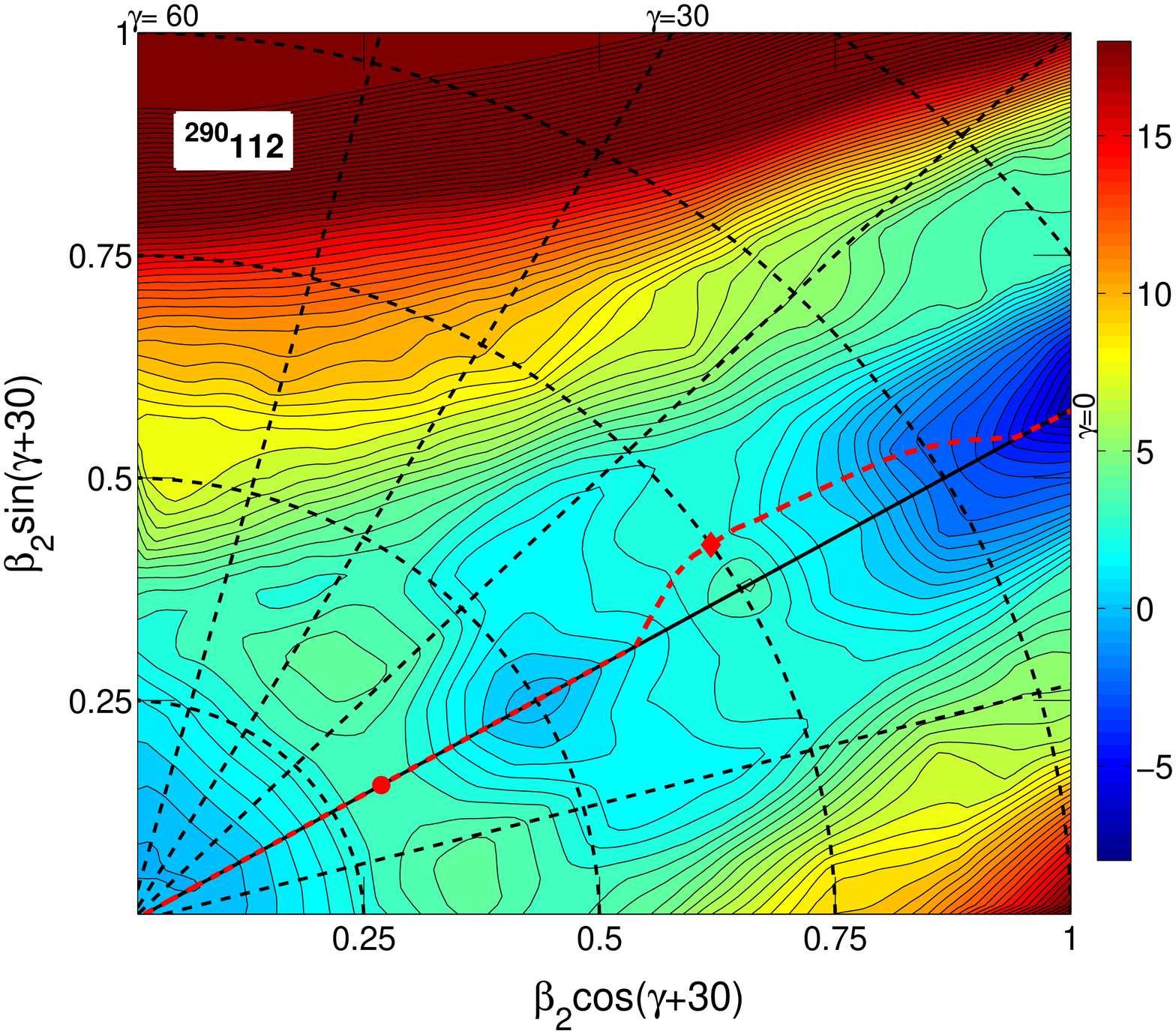}
\hspace{-0.2cm}
\includegraphics[width=7.0cm,angle=0]{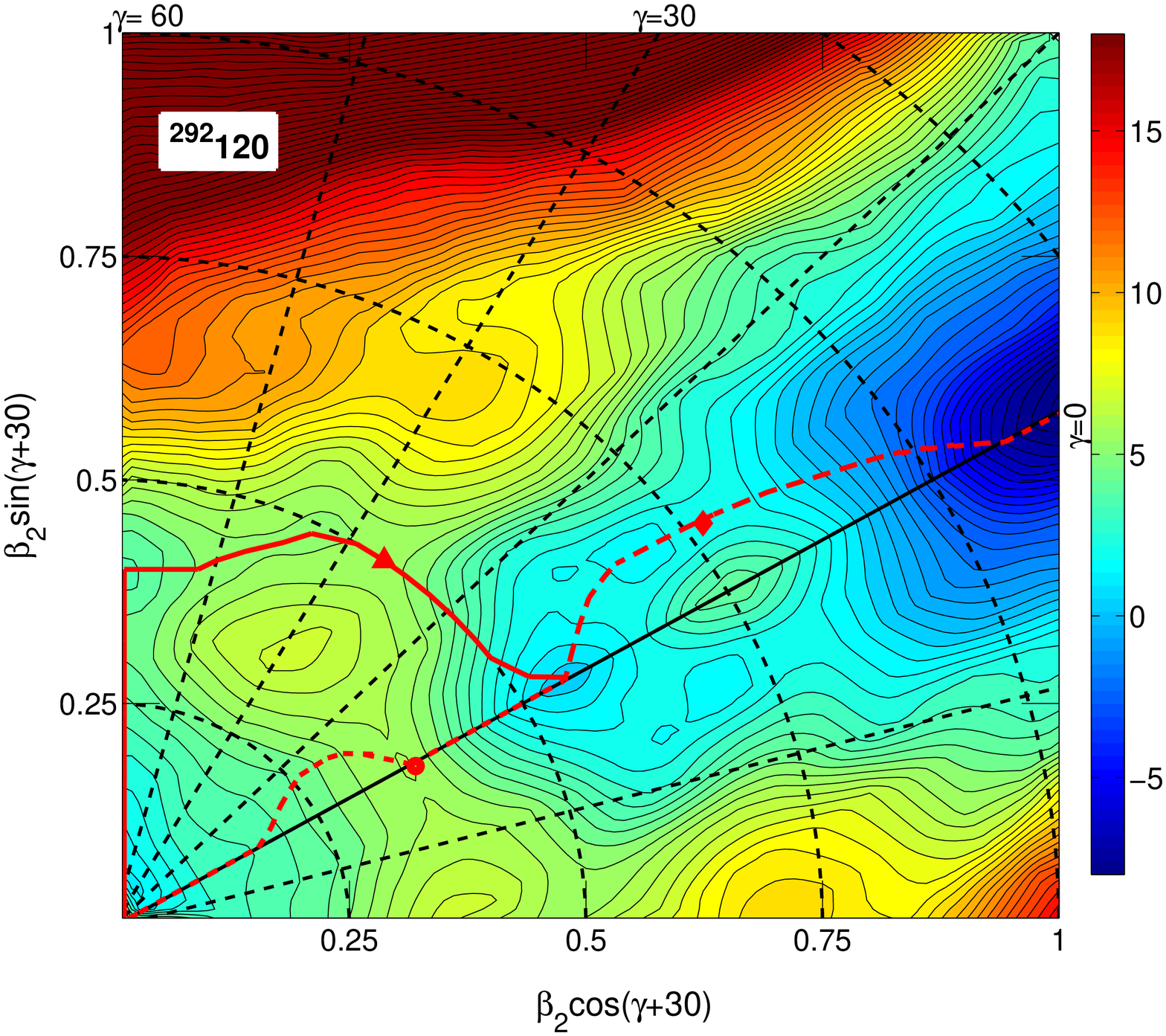}
\end{center}
\vspace{-0.5cm}
\caption{Potential energy surfaces of selected nuclei. The energy
difference between two neighboring equipotential lines is equal to
0.5 MeV. The saddles along the 'Ax' and 'Tr-A' fission pathes
are shown by solid circles and triangles, respectively. The 
solid diamonds show the outer fission barrier saddles. From Ref.\ 
\cite{AAR.12}.}
\label{pes-2D-4panel}
\end{figure*}

  The major difference between the models is related to the use
of experimental information on the energies of fission barriers or
fission isomers in the fit of the model parameters. It turns out 
that only CDFT studies of Refs.\ \cite{AAR.10,LZZ.12,PNLV.12} and 
MM studies of Ref.\ \cite{DPB.07} do not rely on  model 
parametrizations fitted to such data. For example, only spherical 
nuclei were used in the fit of the NL3* \cite{NL3*} and PC-PK1 
\cite{PC-PK1} CDFT parametrizations which provide a successful 
extrapolation to fission barriers of actinides \cite{AAR.10,LZZ.12}. 
In addition, the DD-PC1 parametrization fitted to the ground states 
of normal deformed nuclei \cite{DD-PC1} was used successfully in 
the CDFT study of fission barriers in Ref.\ \cite{PNLV.12}.
On the contrary, all successful descriptions of fission barriers in 
actinides within non-relativistic DFT are based on the parametrizations 
which explicitly use either fission barrier heights (SkM* \cite{SkM*} in 
Skyrme DFT and D1S \cite{D1S-a} in Gogny DFT) or fission isomer energies 
(UNEDF1 \cite{SDFT.12} in Skyrme DFT), which are strongly 
correlated with the inner fission barrier heights, in the fit of model 
parameters. It is also necessary to recognize that 
experimental information on fission barriers have been used in the fit 
of the macroscopic part of the MM models. For example, the macroscopic part 
of Ref.\ \cite{MPS.01} used in the calculations of Ref.\ \cite{KJS.10} 
employs $a_s$ and $\kappa_s$ parameters of the surface term adjusted 
to the large heights of experimental fission barriers of relatively light 
nuclei in Ref.\ \cite{MN.81}. The parameters of the finite-range liquid-drop 
model (FRLDM), used in the extensive MM studies of Ref.\ \cite{MSI.09}, 
have also been very carefully fitted to experimental data on fission 
barriers across the nuclear chart in Ref.\ \cite{MSI.04}.

\section{Extrapolation to superheavy nuclei}

\begin{figure*}
\centering
\includegraphics[angle=0,width=13.0cm]{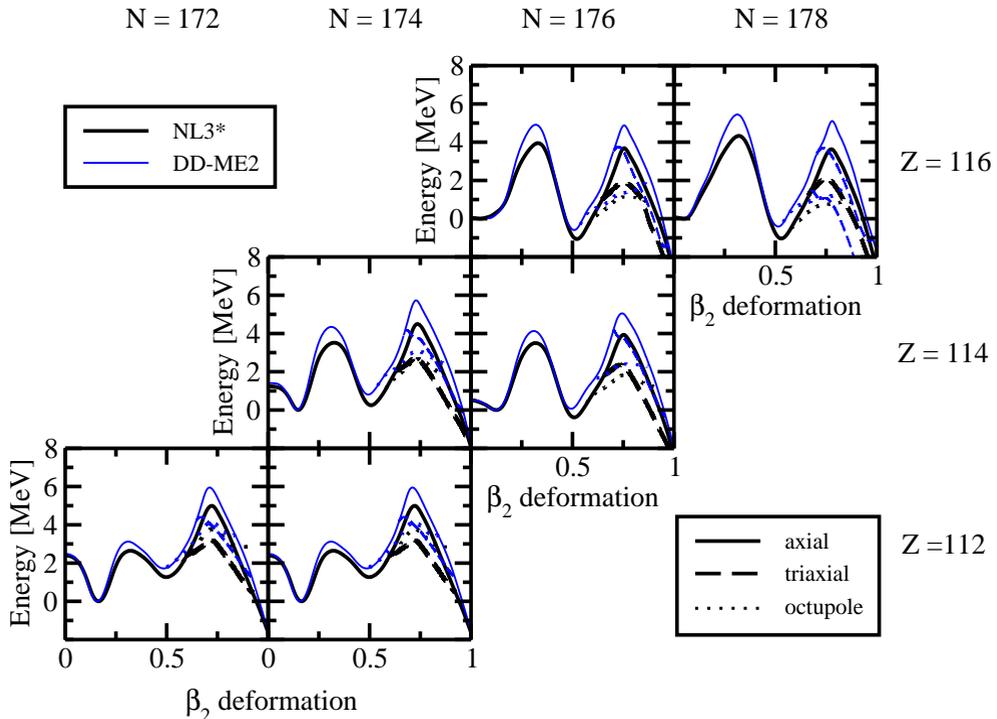}
\vspace{-0.5cm}
\caption{Deformation energy curves for the $Z=112,\,114$ and 116 nuclei 
obtained with the NL3* and DD-ME2 parameterizations. Solid lines correspond 
to axial solutions with reflection symmetry (A), dashed lines to triaxial 
solutions with reflection symmetry (T), and dotted lines to octupole 
deformed solutions with axial symmetry (O). Note that the T and O solutions 
are shown only in the deformation range in which they are lower in energy 
than the axial solution. From Ref.\ \cite{AAR.12}.}
\label{pot3a}
\vspace{-0.5cm}
\end{figure*}

 Fig.\ \ref{pes-2D-4panel} shows that the topology of potential energy
surfaces (PES) of superheavy nuclei is much more complex than the one 
in actinides (see Fig.\ 6 in Ref.\ \cite{AAR.10}) and that it varies 
drastically with the change of particle numbers. The gross structure 
of these PES's is defined by the fact that the total energy is generally 
increasing when moving away from the $\gamma=0^{\circ}$ axis; so it looks 
like a canyon. However, there are local structures inside the canyon 
(such as two triaxial hills with  moderate deformations ($\beta_2 \sim 
0.35,\,\,\gamma \sim \pm30^{\circ}$)  which define the variety of fission 
pathes in the region of inner fission barrier. The axially symmetric 
fission path (labelled as 'Ax') between the normal-deformed/spherical minimum 
and the superdeformed minimum at $\beta_2 \sim 0.5-0.6$ is visible in the 
$^{290}112$ nucleus. As seen in the $^{292}$120 nucleus, 
the $\gamma$-softness of the PES can lead to the development of 
$\gamma$-deformation along the shoulder of the inner fission barrier 
without affecting the saddle point. However, in some nuclei this 
$\gamma-$softness of the PES leads to moderate $\gamma$-deformations
($\gamma \sim 10^{\circ}$) at the saddle; then the fission path is called 
'Ax-Tr'.  When two triaxial hills with ($\beta_2 \sim 0.35,\,\,\gamma 
\sim \pm30^{\circ}$) separate from the walls of the PES canyon, a 
valley between hills and the walls is formed along which the fission 
path 'Tr-A' proceeds. The lowest in energy saddles of these fission 
pathes are shown below in Fig.\ \ref{FB-SHE}.

  The presence of a doubly-humped fission barrier structure in SHN 
seen in Fig.\ \ref{pot3a} is an example of the most striking difference 
between the relativistic and non-relativitis calculations; no
outer fission barrier appears in absolute majority of non-relativistic 
calculations in the $Z\geq 110$ SHN. One can  see that the inclusion of 
triaxiality or octupole deformation always  lowers (by around 2 MeV in 
the majority of the nuclei) the outer fission barrier. The underlying 
shell structure clearly defines which of the saddle points (triaxial 
or octupole deformed) is lower in energy. For example, the lowest saddle 
point is obtained in triaxial calculations in proton-rich nuclei with 
$N < 174$ (Ref.\ \cite{AAR.12}). On the contrary, the lowest saddle 
point is obtained in octupole deformed calculations in neutron-rich 
nuclei with $N > 174$.

\begin{figure}[ht]
\centering
\includegraphics[width=8.0cm,clip]{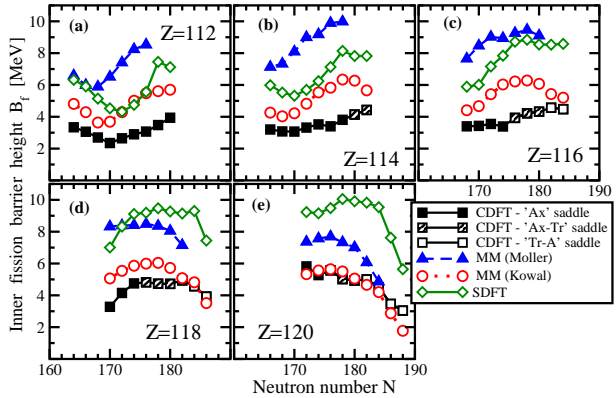}
\vspace{-0.0cm}
\caption{Inner fission barrier heights $B_f$ as a function
of neutron number $N$. The results of the MM calculations
are taken from Ref.\ \cite{MSI.09} (labeled as 'MM (M{\"o}ller)')
and Ref.\ \cite{KJS.10} (labeled as 'MM (Kowal)'). The results
of the Skyrme DFT calculations with the SkM* parametrization
(labelled as 'SDFT') are taken from Ref.\ \cite{SBN.13}. The position
of the inner fission barrier saddle in deformation space varies as
a function of particle number. In the case of the CDFT results
obtained with the NL3* parametrization, the labeling of Ref.\ \cite{AAR.12} 
is used in order to indicate  whether the saddle is axial (labeled as 
'Ax'), has small ($\gamma \sim 10^{\circ}$, labeled as 'Ax-Tr') or large 
($\gamma \sim 25^{\circ}$, labeled as 'Tr-A') $\gamma-$deformations 
in the RMF+BCS calculations.}
\label{FB-SHE}
\end{figure}


  Fig.\ \ref{FB-SHE} shows how the models which have been 
benchmarked in a systematic way in the actinides (see Fig.\ 
\ref{N-dep}) extrapolate to the region of superheavy nuclei.
Note that the results of the MM calculations of Ref.\ 
\cite{KJS.10} labeled as 'MM(Kowal)' are not shown on 
Fig.\ \ref{N-dep}. However, they describe inner fission barriers 
of actinides very accurately. One can see that the model 
predictions vary wildly; the difference in inner fission barrier 
heights between different models reaches 6 MeV in some nuclei. 
This is despite the fact that these models describe the inner
fission barriers in actinides with a comparable level of accuracy. 
The more surprising fact is that the prediction of two MM models 
differ so substantially; in reality the 'MM (Kowal)' model 
predictions are closer to the CDFT ones than to the 'MM (M{\"o}ller)' 
predictions. 

  Let us mention a few possible sources for the differences in the 
predictions of different models. The different location of the 
``magic'' shell gaps in superheavy nuclei in the 
macroscopic+microscopic model (at $Z=114$, $N=184$),
in Skyrme DFT (predominantly at $Z=126$, $N=184$) and in CDFT (at 
$Z=120$, $N=172$) results in different single-particle 
structures at the deformations typical for the ground states and
the saddles of the inner fission barriers. These differences will
definitely affect the inner fission barriers. The sensitivity of
fission barriers to the underlying shell structure can be illustrated
by the reduction of the inner fission barrier height of actinides 
due to triaxiality (Ref.\ \cite{AAR.10}) [or outer fission barrier 
height due to octupole deformation (Ref.\ \cite{MSI.09})] caused by the 
level densities in the vicinity of the Fermi surface which are lower 
at triaxial [octupole] shape as compared with axial one. From our 
point of view, this represents one of the most dominant sources for
the differences in the predictions of the variuos classes of models. 
However, it does not explain the differences within one 
class of models seen, for example, in  the 'MM (Kowal)' and 
'MM (M{\"o}ller)' results of Fig.\ \ref{FB-SHE}.

 Different pairing schemes are used in the calculations of fission
barriers in SHN (see Table IV in Ref.\ \cite{AAR.12}
for review). This is another potential source of errors in model
calculations, which according to Ref.\ \cite{KALR.10} can reach up to 
1 MeV. One should note that particle number projection is ignored 
in the absolute majority of fission barrier calculations. However,
the analysis of rotational structures in actinides in Ref.\ 
\cite{AA.13} leads to the conclusion that approximate particle number 
projection by means of the Lipkin-Nogami method decrease the theoretical 
error bar in the description of the moments of inertia as compared
with unprojected calculations. A similar situation maybe expected also for 
fission barriers. The difference in the effective mass of the nucleon at 
the Fermi surface does not seem to play an important role (see discussion
in Ref.\ \cite{AAR.12-int}. However, the deficiencies in the treatment 
of rotational and vibrational corrections (see Refs.\ \cite{BHB.04}) 
may be responsible for some differences between the various model calculations.

\section{Conclusions}

   Covariant density functional theory provides a good description 
of experimental data on fission barriers in actinides. Contrary to 
the majority of other models, the fit of these relativistic 
parametrizations does not include experimental data on fission 
barriers or isomers. Nonetheless good agreement with experiment
is achieved in this region. This fact enhances the reliability
of CDFT. The similarity of the description of fission barriers 
achieved by different classes of the theoretical models in actinides
does not translate into the similarities of the predictions for
superheavy nuclei. This represents a major theoretical challenge
for the future.

 This work has been supported by the U.S. Department of Energy 
under the grant DE-FG02-07ER41459 and by the DFG cluster of 
excellence ``Origin and Structure of the Universe''.


\end{document}